\documentclass[twocolumn,12pt]{revtex4}
\usepackage{mathtools}
\usepackage{amsmath,bbm}
\usepackage{amssymb,bm}
\usepackage{array}
\usepackage{graphicx}
\usepackage{enumerate}
\usepackage{slashed}
\begin{document}
\title{$\psi(2S)$ enhancement in p$-$Pb collision as an indication of QGP formation at the LHC}

\author{S. Ganesh\footnote{Corresponding author:\\Email: gans.phy@gmail.com}}

\author{Captain R. Singh}

\author{M. Mishra}

\affiliation{Department of Physics, Birla Institute of Technology and Science, Pilani - 333031, INDIA}
\begin{abstract} 
Proton-nucleus collisions serve as an important baseline for the understanding and interpretation of the nucleus-nucleus collisions. These collisions have been employed to characterize the cold nuclear matter effects at SPS and RHIC energies for the past several years, as it was thought that Quark-Gluon Plasma (QGP) is not formed in such collisions. However, at the Large Hadron Collider (LHC), there seems a possibility that QGP is formed during proton-lead (p$-$Pb) collisions. In this work, we have derived an expression for gluon induced excitation of $J/\psi$ to $\psi(2S)$, using pNRQCD, and show that the relative enhancement of $\psi(2S)$ vis a vis $J/\psi$, especially at high $p_T$, gives further indication that the QGP is indeed formed in p$-$Pb collisions at the most central collisions at LHC energy. $J/\psi$ and $\psi(2S)$ suppression effects seen at ALICE are also qualitatively explained.
\end{abstract}
\maketitle
\section{Introduction}
   Quark-Gluon Plasma (QGP) is a deconfined state of quarks and gluons, and is currently a subject of much theoretical and experimental research. QGP is produced in heavy-ion collisions, e.g., Au$-$Au collisions at Relativistic Heavy-ion Collider (RHIC) or Pb$-$Pb collisions at the Large Hadron Collider (LHC) experiments. Some of the signatures of QGP include heavy quarkonium ($J/\psi$ or $\Upsilon$) suppression, collective flow and photon/dilepton production etc. Usually, heavy quarkonium is suppressed to a much larger extent, if QGP is formed~\cite{mats, Chu, Madhu1, gans1, gans2, rishi, Madhu2} in the heavy-ion collision experiments. However, at the LHC, $J/\psi$ yield may actually increase due to recombination~\cite{gans3, stathadref, twice1ref, trans2ref}, which would obviously reduce the effective suppression.
   For proton-Lead (p$-$Pb) collisions, the heavy quarkonium yield may be explained using Cold Nuclear Matter (CNM) effects itself~\cite{medvel}. However, these do not necessarily prove beyond reasonable doubt that QGP is not formed in p$-$Pb collisions. 
There have also been attempts to explain the p$-$Pb data using hot nuclear matter effects~\cite{refA1, refA2, refA3, refA4}. 
In this work, we attempt to explore the yield enhancement of the charmonium state $\psi(2S)$ w.r.t. $J/\psi$, at high $p_T$, as a possible indication of the presence of QGP. One possible explanation for yield enhancement is secondary recombination of $c$ and $\bar{c}$ pair. However, it is unlikely to be a reason for $\psi(2S)$ enhancement at high $p_T$ in the case of p$-$Pb collision. This is because, both theoretical prediction and experimental data indicate that recombination decreases at high $p_T$~\cite{gans3, recomb}. Furthermore, secondary recombination depends quadratically on the number of $c$ and $\bar{c}$ pairs~\cite{thews, gans3, rituraj}, which would be very less for p$-$Pb collision.
   We argue that a possible reason for $\psi(2S)$ enhancement could be due to the gluon induced excitation of $J/\psi(1S)$ to $\psi(2S)$. We further argue that in a medium of equilibrated gluon distribution, this gluon induced excitation increases with $p_T$ of $J/\psi$. This can be understood in the following way. When the gluon medium achieves equilibrium, it would follow the Bose-Einstein distribution, which results in the concentration of gluons in the low energy regime. The gluon density would then decrease exponentially with gluon energy, $E_g$. 
The mass difference between $\psi(2S)$ and $J/\psi(1S)$, is about $0.6$ GeV. The energy equivalent to it needs to be provided by the gluon. But in a Bose-Einstein distribution, $\frac{1}{\exp(Eg/T) - 1}$, 
most of the gluons would have much smaller energy than $0.6$ GeV. However, for $J/\psi$ with large $p_T$, there would be a blue-shift in the gluon energy in the $J/\psi$ frame of reference, in the forward direction, and a red-shift in the backward direction. 
We represent this Doppler shift as $D(v_{rel}, \theta) = \gamma \frac{E_g}{T}(1 - v_{rel}\cos(\theta))$, where $v_{rel}$ is the relative velocity between the medium and $J/\psi$, and $\theta$ is the angle between $v_{rel}$ and incoming gluon momentum. The subsequent Bose-Einstein distribution, then becomes, 
\begin{equation}
f_g(E_g,v_{rel},\theta) = \frac{1}{\exp(D(v_{rel},\theta)) - 1}. 
\end{equation}
The modified Bose-Einstein distribution, $f_g(E_g,v_{rel},\theta)$, leads to an exponential increase in the availability of gluons with energy around $0.6$ GeV, leading to a significant number of $J/\psi$ getting excited to the $\psi(2S)$ state. The above effect is elaborated more in Sec.~\ref{sec:ATLAScomp} (see Fig.~\ref{fig:doppler}). 
As a side note, the mass difference of $0.6$ GeV is expected to decrease with temperature.
Thus, one would qualitatively expect that the gluon induced enhancement from $J/\psi(1S)$ to $\psi(2S)$ would increase with $p_T$. We explore this analytically in the framework of pNRQCD, and compare the results with the preliminary ATLAS data~\cite{ATLAS, new_ATLAS} at $5.02$ TeV. 
The $p_T$ values of $J/\psi$ and $\psi(2S)$ at ATLAS~\cite{ATLAS, new_ATLAS} are high. 
We perform the analysis in the rest frame of the $J/\psi$. In the rest frame, the gluons of interest have an energy of around $0.6$ GeV (mass difference between $J/\psi$ and $\psi(2S)$) after the blue shift.  
The population of higher energy gluons decrease with increasing energy.
Since the gluons of interest are ultra soft gluons even in the $J/\psi$ rest frame, it allows us to analyze the phenomenon within the framework of pNRQCD.  
pNRQCD as an effective field theory, was initially proposed in~\cite{pineda2}. A good overview on pNRQCD has been given in~\cite{pineda,nora3}.

The suppression effects are also included in the present work.
For an apples to apples comparison of the gluon induced dissociation with the gluon induced enhancement presented in this work, we utilize the model of gluon induced dissociation, developed in~\cite{Nendzig}, using the same pNRQCD Lagrangian. We calculated suppression of $\psi(2S)$ and $J/\psi$, and compare it with ALICE data~\cite{alicepsi}.
A crucial aspect is that the binding energy of $\psi(2S)$ is much smaller than the energy gap between $J/\psi$ and $\psi(2S)$. This is expected to result in the $\psi(2S)$ dissociation to be significantly higher than the $J/\psi$ to $\psi(2S)$ excitation. However, the relatively small finite QGP size (lifetime) in a p$-$Pb collision, imposes significant restriction on the $\psi(2S)$ dissociation, especially at high $p_T$. The dissociation process needs to get completed within the QGP phase itself, and not carry over to the hadronic phase. In the hadronic phase, the intermediate octet state (after absorption of a gluon) is much more likely to emit a gluon and form a bound state rather than dissociate into naked $c$ and $\bar{c}$. We revisit this again in Sec.~\ref{sec:Results}.
This causes the dissociation rate to reduce.

Seminal work on heavy quark bound states and their interaction with gluon was done by Peskin~\cite{Pes1} and Bhanot and Peskin~\cite{Pes2}. Subsequent work on heavy quark bound states can be found in~\cite{ghig, nora1, nora2, Nendzig}. 

  The organization of the rest of the article is as follows. In Sec.~\ref{sec:formulation}, the $1S \rightarrow 2S$ transition cross-section, $\sigma$, and the excitation rate, $\Gamma_{1S\rightarrow 2S}$, are calculated. The dissociation processes are outlined in Sec.~\ref{sec:diss}. This is followed by Sec.~\ref{sec:Results}, where the results are shown and compared with both ATLAS and ALICE experimental data. Finally, in Sec.~\ref{sec:conclusion}, we draw our final conclusions. 

\section{{\MakeLowercase p}NRQCD formulation}
\label{sec:formulation}
\subsection{Derivation of the cross-section}
\begin{figure}[h!]
\includegraphics[width=80mm,height=80mm]{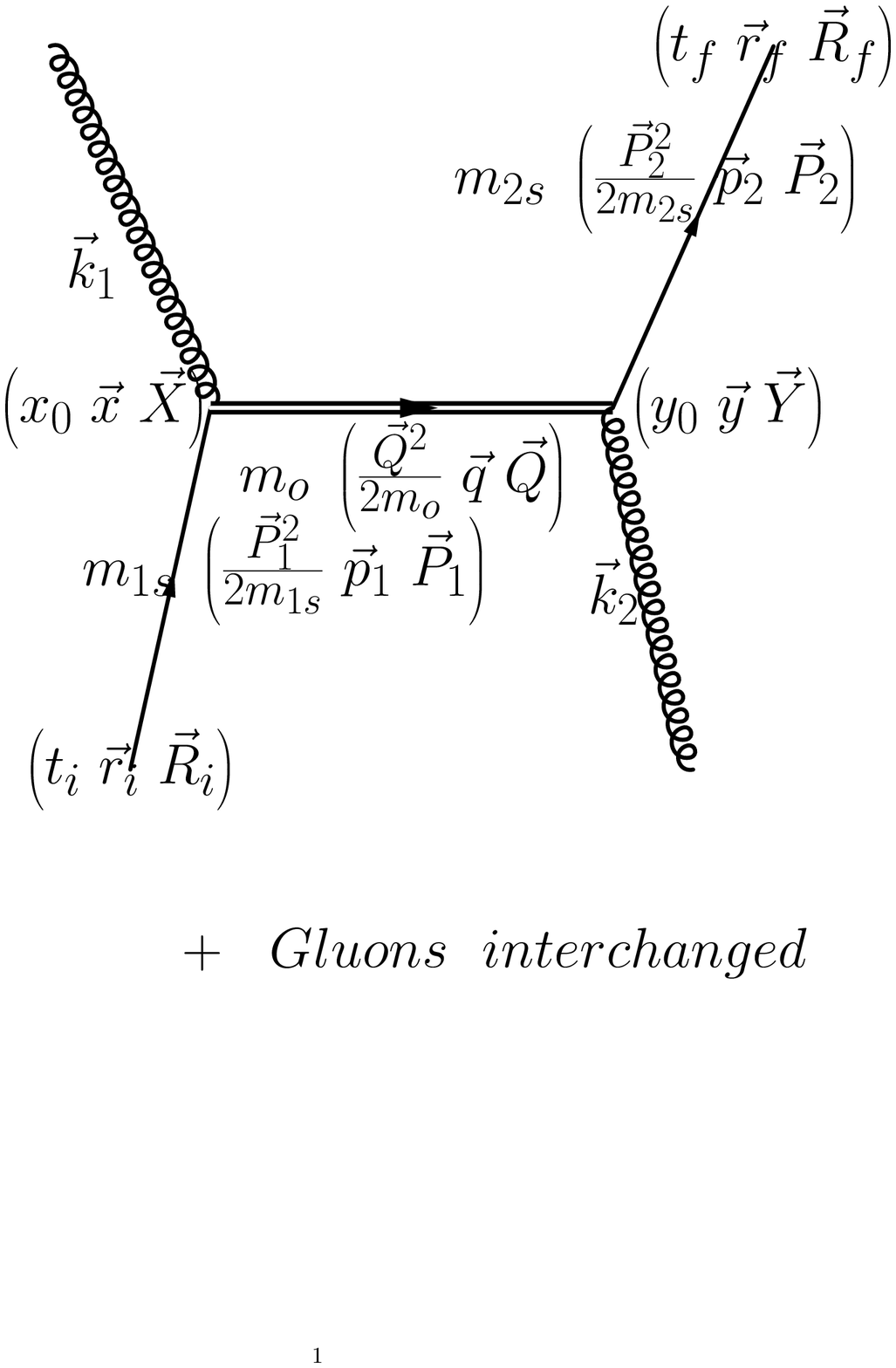}
\caption{Feynman diagrams for gluon induced excitation of $1S$ state to $2S$ state. The second diagram is just the first diagram with gluons interchanged.}
\label{fig:1S_2S}
\end{figure}

 In this section, the cross-section $\sigma$ is calculated using the potential non-relativistic perturbative QCD (pNRQCD) formulation. The corresponding excitation rate $\Gamma_{1S \rightarrow 2S}$ is calculated in the sub section~\ref{sec:rate_constant}, at the end of the current section. The relevant part of the pNRQCD Lagrangian used for the calculation is given by
\begin{math}
\nonumber g Tr \left ( S^{\dagger}(\vec{r}.\vec{E})O + O^{\dagger}(\vec{r}.\vec{E})S \right ).
\end{math}
The variables, $S$, $O$, $E$ and $g$, refer to the singlet, octet, gluonic chromo-electric field and coupling constant, respectively. 
Figure~\ref{fig:1S_2S} depicts the Feynman diagrams for the processes involved. 
The net amplitude would be the sum of the two diagrams shown in Fig.~\ref{fig:1S_2S}. In the second diagram, the incoming singlet first emits an outgoing gluon of $3$-momentum $\vec{k}_2$, and subsequently absorbs a gluon with $3$-momentum $\vec{k}_1$. 
The possible singlet states, which could significantly contribute to $\psi(2S)$ via gluo-excitation can be $J/\psi$, $\eta_c$ or $\chi(1P)$. We mainly focus on $J/\psi$. The case, when the incoming singlet is either $\eta_c$ or $\chi(1P)$, is discussed in the latter part of this section. 
\label{sec:NRQCD}

The notation and variables used are described as follows. The center-of-mass coordinates are: 
\begin{itemize}
\item $\vec{R}_i = (\vec{x}_{qi} + \vec{x}_{\bar{q}i})/2$; ~~$\vec{R}_f = (\vec{x}_{qf} + \vec{x}_{\bar{q}f})/2$.
\end{itemize}
The relative motion (RM) coordinates are:
\begin{itemize}
\item $\vec{r}_i = (\vec{x}_{qi} - \vec{x}_{\bar{q}i})$; ~~~~~~$\vec{r}_f = (\vec{x}_{qf} - \vec{x}_{\bar{q}f})$.
\end{itemize}

The singlet and octet fields are $S = S_{nlm}\,I_3/\sqrt{N_c}$ and $O_q = \sqrt{2} O_q^bT^b$, with $\frac{q^2}{m_c}$ being the energy eigenvalues of the octet state, and $T^b$ being the generators of $SU(N_c)$. $m_c$ is the mass of the charm quark and anti-quark.
In particular, we denote the singlet $1S$ and $2S$ wavefunctions as $S_{1S}(\vec{x})$ and $S_{2S}(\vec{x})$, respectively. 
$\vec{P}_1$, $\vec{P}_2$, $\vec{Q}$, $\vec{k}_1$ and $\vec{k}_2$ refer to the $3$-momentum of the incoming $J/\psi$, outgoing $\psi(2S)$, octet, incoming gluon and outgoing gluon, respectively. The variables, $m_{1S}$, $m_{2S}$ and $m_o$, are the invariant masses of the $J/\psi$, $\psi(2S)$ and octet states, respectively, while $\vec{p}_1$, $\vec{p}_2$ and $\vec{q}$ refer to the relative $3$-momentum between the $c\bar{c}$ pair comprising these particles. $k_{0x}$ is the energy corresponding to $\vec{k}_x$, with $x=1,2$.


We model the process at leading order (LO) in pNRQCD, in a manner similar to~\cite{Nendzig}.
The RM octet propagator at LO would be 
\begin{eqnarray}
\nonumber P^{rm}_O(\vec{x},x_0,\vec{y},y_0)\\
\nonumber = \sum_{qlm} O^{b}_{qlm}(\vec{y})O^{* b'}_{qlm}(\vec{x})\delta^{bb'}e^{-i\frac{q^2}{m_c}(y_0 - x_0)}\\
\rightarrow \sum_{lm}\int dq O^b_{qlm}(\vec{y})O^{* b'}_{qlm}(\vec{x})\delta^{bb'}e^{-i\frac{q^2}{m_c}(y_0 - x_0)}, 
\end{eqnarray}
as the octet states, represented by $q$ is a continuum of states. The octet wavefunctions are normalized to the Dirac delta function.
The RM singlet propagator would be 
\begin{eqnarray}
\nonumber P^{rm}_S(\vec{x},x_0,\vec{r}_i,t_i) = \\
\sum_{nlm} S_{nlm}(\vec{x})S^{*}_{nlm}(\vec{r}_i)e^{-iE_{S}(x_0 - t_i)}
\end{eqnarray}
The variable $E_{S}$ represents the singlet energy eigenvalues.
The center-of-mass (CM) octet propagator would be 
\begin{eqnarray}
\nonumber P^{cm}_{O}(\vec{Y},y_0,\vec{X},x_0) = \\
\int \frac{dE_v}{2\pi} \int \frac{d^3Q}{(2\pi)^3} e^{-i\left ( E_v(y_0 - x_0) + \vec{Q}.(\vec{Y}-\vec{X})\right )}.
\end{eqnarray}
Here, $E_v$ is the center of mass energy of the virtual, off-shell, octet state.
The center-of-mass propagator for the incoming singlet would be 
\begin{eqnarray}
\nonumber P^{cm}_{S}(\vec{X},x_0,\vec{R}_i,t_i) = \\
\int \frac{d^3P_1}{(2\pi)^3} e^{-i \left ( \frac{\vec{P}_1^2}{2m_s}(x_0 - t_i) + \vec{P}_1.(\vec{X}-\vec{R}_i) \right ) }.
\end{eqnarray}
Similarly, let $P^{cm}_{S}(\vec{R}_f,t_f,\vec{Y},y_0)$ be the center-of-mass propagator for the outgoing singlet.
The overall amplitude from $(\vec{r}_i,t_i)$ to $(\vec{r}_f,t_f)$, including the vertex factors, would be 
\begin{eqnarray}
\nonumber G(r_f, t_f, R_f, r_i, t_i, R_i) = \\
\nonumber g^2 C \int dx_0 \int dy_0 \int d^3X \int d^3 Y \int d^3x \int d^3y \\
\nonumber \times \Big \{ P^{cm}_S(\vec{R}_f,t_f,\vec{Y},y_0)P^{rm}_S(\vec{r}_f,t_f,\vec{y},y_0)(\vec{y}.\vec{E}^{a*}_2)\\
\nonumber \times P^{cm}_O(\vec{Y},y_0,\vec{X},x_0)P^{rm}_O(\vec{y},y_0,\vec{x},x_0)(\vec{x}.\vec{E}^a_1)\\
\nonumber \times P^{cm}_S(\vec{X},x_0,\vec{R}_i,t_i)P^{rm}_S(\vec{x},x_0,\vec{r}_i,t_i) \Big \}\\
+ {gluons~interchanged~terms,~~}
\end{eqnarray}
where, 
\begin{eqnarray}
\nonumber C = Tr\left[ \frac{I_3}{\sqrt{N_c}}T^a\sqrt{2}T^b\right] 
\times  Tr\left[ \frac{I_3}{\sqrt{N_c}}\sqrt{2}T^b T^c\right]\\ 
\nonumber = \frac{\delta^{ac}}{2N_c}.
\end{eqnarray}
The superscripts "a" and "c" refer to the species of the incoming and outgoing gluons.
The $T$ matrix for the $1S$ to $2S$ transition is then given by
\begin{eqnarray}
\nonumber T(1S \rightarrow 2S) = \int d^3 r_i \int d^3 r_f e^{-i\left ( \vec{P}_2.R_f\right )}\\
S_{2S}^*(r_f)G(r_f, t_f, R_f, r_i, t_i, R_i) S_{1S}(r_i)e^{i\left ( \vec{P}_1.R_i\right )}.
\end{eqnarray}

The term $\vec{x}.\vec{E}^a_1$ evaluates to $k_{01}(\vec{x}.\hat{\epsilon}_1)e^{-i\left ( \vec{k}_1.\vec{X} + k_{01}x_0 \right )}$, where $\hat{\epsilon}_1$ is the polarization of the incoming gluon.
Similarly, for the outgoing gluon, $\vec{y}.\vec{E}^{a*}_2$ evaluates to $k_{02}(\vec{y}.\hat{\epsilon}^*_2)e^{i\left (\vec{k}_2.\vec{Y} + k_{02}y_0 \right )}$.

We also normalize the singlet wavefunctions as:
\begin{eqnarray}
\nonumber S_{n'l'm'}(\vec{x})S^*_{nlm}(\vec{r_i}) = \delta^3(\vec{r}_i - \vec{x})\delta_{n,n'}\delta_{l,l'}\delta_{m,m'}\\
S_{n'l'm'}(\vec{r_f})S^*_{nlm}(\vec{y}) = \delta^3(\vec{y} - \vec{r}_f)\delta_{n,n'}\delta_{l,l'}\delta_{m,m'}
\end{eqnarray}
where, we have dropped the superscript "b", and taken the wavefunction $O^b_{qlm}(\vec{x}) = O_{qlm}(\vec{x})$.
With all these, 
\begin{eqnarray}
\nonumber T(1S \rightarrow 2S) =  \Bigg [ g^2 C \int \frac{dE_v}{2\pi}\\
\nonumber \sum_{lm} \int dq \langle S_{2S}|(\vec{y}.\hat{\epsilon}^*_2)|O_{qlm} \rangle \langle O_{qlm}|(\vec{x}.\hat{\epsilon}_1)|S_{1S} \rangle \\
\nonumber \times \int d^3X \int d^3Y \int dx_0 \int dy_0\\
\nonumber \times  e^{-i \left ( \vec{P}_2.(\vec{R}_f - \vec{Y}) + \frac{\vec{P}_2^2}{2m_{2S}}(t_f - y_0) + E_{1S}(t_f - y_0) \right ) }\\
\nonumber \times e^{-i \left ( \vec{Q}.(\vec{Y}- \vec{X}) + E_v(y_o - x_o) + \frac{q^2}{m_c}(y_o - x_o) \right )}\\
\nonumber \times e^{-i \left ( \vec{P}_1.(\vec{X}- \vec{R}_i) + \frac{\vec{P}_1^2}{2m_{1S}}(x_o - t_i) + E_{2S}(x_o - t_i) \right )}\\
\nonumber \times e^{-i\left ( \vec{k}_1.\vec{X} + k_{01}x_0 \right )} e^{i\left (\vec{k}_2.\vec{Y} + k_{02}y_0 \right )} \Bigg ]\\
+ {gluons~interchanged~terms,~~}
\end{eqnarray}
where,
\begin{eqnarray}
\nonumber \langle O_{qlm}|(\vec{x}.\hat{\epsilon}_1)|S_{1S} \rangle =\\
\nonumber \int d^3x  O^*_{qlm}(\vec{x})(\vec{x}.\hat{\epsilon}_1)S_{1S}(\vec{x}),\\
\nonumber and~~~~~~~~~~~~~~~~~~~~~~~~~~~~~~~~~~~~~~~~~~~\\
\nonumber \langle S_{2S}|(\vec{y}.\hat{\epsilon}^*_2)|O_{qlm} \rangle =\\
\int d^3y  S^*_{2S}(\vec{y})(\vec{y}.\hat{\epsilon}^*_2)O_{qlm}(\vec{y}).
\end{eqnarray}
The variables $E_{1S}$ and $E_{2S}$ are the energy eigenvalues for the input $J/\psi$ and output $\psi(2S)$ singlet states respectively.
Finally, performing all the integrals: 
\begin{eqnarray}
\label{eq:tmatrix}
\nonumber T(1S \rightarrow 2S) = g^2 C (2\pi) k_{01}k_{02} \int \frac{dE_v}{2\pi} \sum_{lm}\\
\nonumber \Bigg [ \Big \{ \langle S_{2S}|(\vec{y}.\hat{\epsilon}^*_2)|O_{q1,lm} \rangle \langle O_{q1,lm}|(\vec{x}.\hat{\epsilon}_1)|S_{1S} \rangle \\
\nonumber  \times  \frac{1}{2}{\sqrt \frac{m_c}{\frac{\vec{P}_1^2}{2m_{1S}} - E_v + k_{01} + E_{1S}} } \Big \} \\
\nonumber + \Big \{ \langle S_{2S}|(\vec{y}.\hat{\epsilon}_1)|O_{q2,lm} \rangle \langle O_{q2,lm}|(\vec{x}.\hat{\epsilon}^*_2)|S_{1S} \rangle \\
\nonumber  \times \frac{1}{2}{\sqrt \frac{m_c}{\frac{\vec{P}_1^2}{2m_{1S}} - E_v - k_{02} + E_{1S}} } \Big \} \Bigg ]\\
\nonumber \times (2\pi)^3\delta^3(\vec{P}_1 + \vec{k}_1 - \vec{P}_2 - \vec{k}_2) e^{-i\phi}\\
\nonumber \times (2\pi)\delta( \frac{\vec{P}_1^2}{2m_{1S}} + k_{01} - \frac{\vec{P}_2^2}{2m_{2S}} - k_{02} - \Delta m),\\ 
\end{eqnarray}
where, $\phi$ is an arbitrary phase factor $= \vec{P}_2.\vec{R}_f - \vec{P}_1.\vec{R}_i + (\frac{\vec{P}_2^2}{2m_{2S}} + E_{2S})t_f - (\frac{\vec{P}_1^2}{2m_{1S}} + E_{1S})t_i $, and $\Delta m = E_{2S} - E_{1S} (= m_{2S} - m_{1S})$. 
The phase factor does not appear in the final expression for the cross-section.
Finally, the values of $q1$ and $q2$ appearing in Eq.~\ref{eq:tmatrix} are given by:
\begin{eqnarray}
\nonumber \frac{q1^2}{2m_o}=  \frac{\vec{P}_1^2}{2m_{1S}} - E_v + k_{01} + E_{1S}\\ 
\nonumber \frac{q2^2}{2m_o}=  \frac{\vec{P}_1^2}{2m_{1S}} - E_v - k_{02} + E_{1S}.
\end{eqnarray}
The values of $E_v$ used in simulation correspond to varying the octet energy eigenvalues, $\frac{q_1^2}{2m_o}$ and $\frac{q_2^2}{2m_o}$ from $0.1$ GeV to infinity (represented by $100$ GeV). 
In the $J/\psi$ rest frame, $\vec{P}_1 = 0$. Further, one can also ignore $\frac{P_2^2}{2m_{2S}}$, which simplifies the calculations. From the above $T$ matrix, $T(1S \rightarrow 2S)$, we extract out the energy and momentum conserving $\delta$ functions to get the $M$ matrix, 
\begin{eqnarray}
\nonumber M(1S \rightarrow 2S) = 2\pi g^2 C k_{01} k_{02} M_c e^{-i\phi},
\end{eqnarray}
with 
\begin{eqnarray}
\nonumber M_c = \int \frac{dE_v}{2\pi}\\
\nonumber \Bigg \{ \Bigg [ \sum_{lm} \langle S_{2S}|(\vec{y}.\hat{\epsilon}^*_2)|O_{q1,lm} \rangle \langle O_{q1,lm}|(\vec{x}.\hat{\epsilon}_1)|S_{1S} \rangle \\
\nonumber \times \left  ( \frac{1}{2} {\sqrt \frac{m_c}{k_{01} + E_{1S} - E_v} }\right ) \Bigg ]\\
\nonumber + \Bigg [ \sum_{lm}\langle S_{2S}|(\vec{y}.\hat{\epsilon}_1)|O_{q2,lm} \rangle \langle O_{q2,lm}|(\vec{x}.\hat{\epsilon}^*_2)|S_{1S} \rangle \\
\nonumber \times \left  ( \frac{1}{2} {\sqrt \frac{m_c}{-k_{02} + E_{1S} - E_v} }\right ) \Bigg ] \Bigg \}.\\
\end{eqnarray}

We use the $M$ matrix to calculate the cross-section. The average $1S \rightarrow 2S$ transition cross-section, $\sigma$, after dividing by the number of input gluons, is then given by
\begin{eqnarray}
\label{eq:sigma1}
\nonumber \sigma = \frac{1}{2E^T_{1S}2k_1(1 - v_{1S})}\\
\nonumber \times \int \frac{d^3P_2}{(2\pi)^3 2E^T_{2S}}\int \frac{d^3k_2}{(2\pi)^3 2k_{02}} \\
\nonumber  \times C_g \Bigg [ \left ( (2\pi) M_c k_{01}k_{02} \right )^*\left ( (2\pi) M_c k_{01}k_{02} \right )\Bigg ]\\
\nonumber \times (2\pi)^3\delta^3(\vec{P}_2 + \vec{k}_2 - \vec{k}_1) \\
\times (2\pi) \delta \left ( k_{01} - \frac{\vec{P}_2^2}{2m_{2S}} - k_{02} - \Delta m \right),
\end{eqnarray}
where, $C_g = g^2\frac{1}{(2N_c)^2} = 4\pi \alpha \frac{1}{(2N_c)^2}$.
The expression $\delta (k_{01} - \frac{\vec{P}_2^2}{2m_{2S}} - k_{02} - \Delta m)$ in Eq.~\ref{eq:sigma1} gives $k_{02} = k_{01} - \frac{\vec{P}_2^2}{2m_{2S}} - \Delta m \approx k_{01} - \Delta m$.
This value of $k_{02}$ makes $M_c$ independent of both $\vec{P}_2$ and $k_2$, and thus can be taken outside the $d^3P_2$ and $d^3k_2$ integrals. 
It is also understood that $k_2$ = $|\vec{k}_2|$ =$k_{02}$, and similarly $k_1$ = $|\vec{k}_1|$ = $k_{01}$.

In the $J/\psi$ rest frame, $v_{1S} = 0$. We decompose $\vec{k}_2$ and $\vec{P}_2$ into parallel and perpendicular components to $\vec{k}_1$, and approximating $E^T_{2S} = m_{2S} + \frac{P_2^2}{2m_{2S}} \approx m_{2S}$, we get
\begin{eqnarray}
\nonumber \sigma = \frac{C_g(2\pi)}{16m_{1S}{m_{2S}}} k_1 M_c^2 \int d^3k_2 k_2 k_{2\perp}\\
\times \delta \left ( k_1 - \frac{(k_1 - k_{2||})^2 + k_{2\perp}^2}{2m_{2S}} - k_2 - \Delta m \right ).
\end{eqnarray}
Substituting $k_{2\perp} = k_2 \sin(\alpha)$ and $k_{2||} = k_2 \cos(\alpha)$, with $\alpha$ being the angle between $\vec{k}_1$ and $\vec{k}_2$, and evaluating the $\int dk_2$ integral, we obtain:
\begin{eqnarray}
\label{eq:sigma}
\nonumber \sigma = \frac{C_g (2\pi)^2}{16m_{1S}m_{2S}}k_1 M_c^2\\
\times \int_0^{\pi} \left [\sin^2\alpha {\sqrt \frac{m_{2S}}{2\Delta} k_2^4} \right ] d\alpha,
\end{eqnarray}
with $\Delta = \frac{(k_1 \cos(\alpha) - m_{2S})^2}{2m_{2S}} - \left ( k_1 - \frac{k_1^2}{2m_{2S}} - \Delta m \right)$, and $k_2$ is now determined in terms of $k_1$ and others, and is equal to  $\sqrt{2m_{2S} \Delta} + (k_1 cos(\alpha) - m_{2S})$.
One can see that there is a pole when $\Delta \rightarrow 0$. This pole is unphysical, and occurs when $k_1$ is large. In other words, this implies that the above formulation is invalid for very large values of $k_1$. In our simulations, we have limited the value of $k_1$ to $1.0$ GeV.  
Due to the Bose enhancement of the outgoing gluon, we scale the cross-section $\sigma$ in Eq.~\ref{eq:sigma} by the factor $ f_{BE} = (1 + \frac{1}{\exp(k_{02}/T) - 1}) = (1 + \frac{1}{\exp(k_2/T) - 1})$. 
Thus we finally obtain:
\begin{eqnarray}
\nonumber \sigma = \frac{C_g (2\pi)^2}{16m_{1S}m_{2S}}k_1 \\
\times M_c^2 \int_0^{\pi} \left [f_{BE}\sin^2\alpha {\sqrt \frac{m_{2S}}{2\Delta} k_2^4} \right ]d\alpha.
\end{eqnarray}

\subsection{Evaluation of the correlation term}
We now evaluate the correlation term $\sum_{l'm'}\langle S_{2S}|(\vec{y}.\hat{\epsilon}^*_2)|O_{ql'm'}\rangle \langle O_{ql'm'}|(\vec{x}.\hat{\epsilon}_1)|S_{1S}\rangle$. Trivially, $\vec{x}.\hat{\epsilon}_1 = |\vec{x}| (\hat{x}. \hat{\epsilon}_1)$, where $\hat{x}$ is a unit vector along $\vec{x}$. Similarly, $\vec{y}.\hat{\epsilon}^*_2 = |\vec{y}| (\hat{y}.\hat{\epsilon}^*_2)$. We can now separate the radial and angular part of the correlation term.
In general,
\begin{eqnarray}
\nonumber \langle O_{ql'm'}|(\vec{x}.\hat{\epsilon}_1)|S_{nlm} \rangle = \\
\nonumber \int dx\,x^2 O^*_q(|\vec{x}|)\,|\vec{x}|\,S_{n}(|\vec{x}|) \\
\times \int d\Omega Y^*_{l'm'}(\theta,\phi)(\hat{x}.\hat{\epsilon}_1) Y_{lm}(\theta,\phi),
\end{eqnarray}
where $S_n(|\vec{x}|)$ and $O_q(|\vec{x}|)$ are the pure radial part of the singlet and octet wavefunction, respectively.
For the input gluon, we average over the $\epsilon_{+}$ and $\epsilon_{-}$ polarization states of the gluon, giving the angular part as:
\begin{eqnarray}
\label{eq:eps_ave}
\nonumber \frac{1}{2}\int d\Omega \Big \{ Y^*_{l'm'}(\theta,\phi)\\
\times \left (\frac{\sin(\theta)e^{i\phi}}{\sqrt 2} + \frac{\sin(\theta)e^{-i\phi}}{\sqrt 2} \right ) Y_{lm}(\theta,\phi) \Big \},
\end{eqnarray}
where $Y_{lm}(\theta, \phi)$ and  $Y^*_{l'm'}(\theta, \phi)$ are the spherical harmonics. The variables $l$ and $l'$ denote the azimuthal quantum number, while, $m$ and $m'$ denote the magnetic quantum number. The integral $d\Omega$ is over the whole solid angle, defined by the angles $\theta$ and $\phi$.
For the output gluon, we sum over the $\epsilon_{+}$ and $\epsilon_{-}$ polarization states of the gluon giving,
\begin{math}
\nonumber \int d\Omega Y^*_{lm}(\theta,\phi)\left (\frac{\sin(\theta)e^{-i\phi}}{\sqrt 2} + \frac{\sin(\theta)e^{i\phi}}{\sqrt 2} \right ) Y_{l'm'}(\theta,\phi).
\end{math}
Let us analyze the expression a little bit more. For $\epsilon = \epsilon_+$, the angular part evaluates to
\begin{eqnarray}
\label{eq:eps_p}
\int d\Omega Y^*_{l'm'}(\theta,\phi)(\frac{\sin(\theta)e^{i\phi}}{\sqrt 2}) Y_{lm}(\theta,\phi).
\end{eqnarray}
The $d\Omega$ integral evaluates to $1$, if $l' = l+1$, and $m' = m+1$, and $0$, otherwise. Similarly, when $\epsilon = \epsilon_{-}$, the $d\Omega$ integral evaluates to $1$, if  $l' = l+1$, and $m' = m-1$, and $0$, otherwise. This indicates that when the singlet is in $1S$ state i.e. $l=0$, it will transit to an octet $1P$.
On evaluating the other correlation term $\langle S_{2S}|(\vec{y}.\hat{\epsilon}^*_2)|O_q\rangle$, in a similar manner, we find that the octet $1P$ will be converted to singlet $\psi(2S)$.
Putting all this together, the product of the two correlation function evaluates to: 
\begin{eqnarray}
\nonumber \frac{1}{2}\sum_{l'm'} \Big [\int d\Omega Y^*_{00}(\theta,\phi)\\
\nonumber \times \left (\frac{\sin(\theta)e^{-i\phi}}{\sqrt 2} + \frac{\sin(\theta)e^{i\phi}}{\sqrt 2} \right ) Y_{l'm'}(\theta,\phi)\\
\nonumber \times \int d\Omega Y^*_{l'm'}(\theta,\phi)\\
\times \left (\frac{\sin(\theta)e^{i\phi}}{\sqrt 2} + \frac{\sin(\theta)e^{-i\phi}}{\sqrt 2} \right ) Y_{00}(\theta,\phi) \Big ]
\end{eqnarray}
The expression is non-zero only for $l' = 1$. This gives the angular part as:
\begin{eqnarray}
\nonumber \frac{1}{2}\Bigg [ \Bigg ( \int d\Omega \frac{1}{\sqrt{4\pi}}\frac{\sin(\theta)e^{-i\phi}}{\sqrt 2}\frac{3}{\sqrt{8\pi}}\sin(\theta)e^{i\phi} \\
\nonumber \times 
\int d\Omega \frac{3}{\sqrt{8\pi}}\sin(\theta)e^{-i\phi}\frac{\sin(\theta)e^{i\phi}}{\sqrt 2} \frac{1}{\sqrt{4\pi}} \Bigg )\\ 
\nonumber +
\Bigg ( \int d\Omega \frac{1}{\sqrt{4\pi}}\frac{\sin(\theta)e^{i\phi}}{\sqrt 2} \frac{3}{\sqrt{8\pi}}\sin(\theta)e^{-i\phi} \\
\times
\nonumber \int d\Omega \frac{3}{\sqrt{8\pi}}\sin(\theta)e^{i\phi}\frac{\sin(\theta)e^{-i\phi}}{\sqrt 2}\frac{1}{\sqrt{4\pi}}  \Bigg ) \Bigg ]\\~~~~~~~~~~~~~~~~~~~=1.~
\end{eqnarray}

We are now in a position to analyze the cross-section for $\chi_c(1P)$ and $\eta_c$ particles transition to $\psi(2S)$.
\subsection{$\chi(1P)$}
From Eq.~\ref{eq:eps_p} and from properties of spherical harmonics, it can be seen that in general, the expression in Eq.~\ref{eq:eps_p} is of the form
\begin{eqnarray}
\nonumber \delta(l+1-l')\delta(m+1-m')(..) \\
+ \delta(l-1-l')\delta(m+1-m')(..),
\end{eqnarray} 
for the $\epsilon_+$ polarization. 
Similarly,
\begin{eqnarray}
\nonumber \delta(l+1-l')\delta(m-1-m')(..) \\
+ \delta(l-1-l')\delta(m-1-m')(..),
\end{eqnarray}
for the $\epsilon_{-}$ polarization. 

This would imply that when the input singlet is in $1P$ state, the octet would be either in $l=0$ or $l=2$ state. This further implies that the output singlet will have to be in either $l=1$ or $l=3$ state, but certainly not $l=0$ state. As a consequence of this, the cross-section for $\chi(1P)$ to $\psi(2S)$, transition would be zero. 
However, as it may occur through other mechanisms, we expect $\psi(2S)$ production from $\chi(1P)$ to be suppressed.  
\subsection{$\eta_c$}
The particle, $\eta_c$, is a color singlet and spin singlet particle. A chromo-magnetic sector is required to modify the spin state. A chromo-magnetic vertex would be characterized by the operator $\frac{1}{2m_c}\sigma.\vec{B}^a$, where $m_c$ is the charm quark mass, and $\vec{B}^a$ is the chromo-magnetic field.
For a real transverse gluon, $B_z^a$ = 0. This gives the vertex operator as $\frac{1}{2m_c}(\sigma_xB_x^a + \sigma_yB_y^a)$. This operator would convert a spin $0$ singlet state to spin $\pm 1$ state for the octet, depending on the gluon polarization. 
From considerations of conservation of total angular momentum, the octet wavefunction would then be an s-wave (i.e., $l=0$). This can also be seen explicitly by evaluating the angular part of the correlation function $\langle O|\frac{1}{2m_c}\sigma.\vec{B}^a|S \rangle $. This implies that the second vertex involving the outgoing gluon would again be a chromo-magnetic vertex, since a chromo-electric vertex would modify the value of $l$ by $1$. 
Again from considerations of conservation of angular momentum, a chromo-magnetic vertex would require the spin of the outgoing singlet to be $0$. But the spin of $\psi(2S)$ is $1$. Hence to the order $\alpha^2$, a transition from $\eta_c$ to $\psi(2S)$ is not possible.
One could however argue that, if one of the gluons is off-shell, then the longitudinal polarization of the virtual gluon can lead to the creation of $\psi(2S)$ with longitudinal polarization. 
We know from~\cite{bodwin}, that the chromo-magnetic vertex is higher in velocity scale, than the chromo-electric vertex, by a factor of heavy quark velocity $v$.   
Hence, the amplitude for the process $\eta_c \rightarrow octet \rightarrow \psi(2S)$, would be suppressed by a factor $v^2$, and the cross-section by a factor $v^4$.

Due to the above reasons, we ignore the contribution of both $\chi(1P)$  and $\eta_c$ particles to $\psi(2S)$ via gluo-excitation.

\subsection{The rate constant $\Gamma_{1S\rightarrow 2S}$}
\label{sec:rate_constant}
The rate constant, $\Gamma_{1S\rightarrow 2S}$, is obtained by integrating the $1S \rightarrow 2S$ transition cross-section, $\sigma$, with the gluon distribution function, $f_g(E_g,v_{rel},\theta) = \frac{g_d}{e^{\frac{\gamma E_g}{T}(1 - v_{rel} \cos(\theta))} - 1}$, where $g_d$ is the number of gluon degree of freedom $=16$. In this section, we have used $E_g$ in place of $k_1$ (or $k_{01}$) for the incoming gluon energy.
 
In other words,  
\begin{eqnarray}
\label{eq:rateconstant}
\nonumber \Gamma_{1S\rightarrow2S} = \\
\int \frac{1}{4\pi^2} E_g^2 \sin(\theta)f_g(E_g,v_{rel},\theta)\,\sigma\,dE_g d\theta
\end{eqnarray}
\begin{eqnarray}
\nonumber = \frac{1}{4\pi^2}\int \frac{ E_g g_dT\,\sigma}{v_{rel}\gamma}\\ 
\times \ln\Bigg [ \frac{e^{\frac{\gamma E_g}{T}(1 + v_{rel})} - 1}
{e^{\frac{2\gamma E_g v_{rel}}{T}} \left (e^{\frac{\gamma E_g}{T} (1- v_{rel})} - 1\right ) } \Bigg ]dE_g.
\end{eqnarray}
The fraction of the number of $1S$ particles converted to $2S$ is then given by
\begin{math}
\nonumber \Delta n_{1S\rightarrow 2S} =  1 - \exp\left (-\int_{t_0}^{t_{QGP}} \Gamma_{1S\rightarrow2S}\,dt \right ).
\end{math}
The variables $t_0$ and $t_{QGP}$ indicate the thermalization time and lifetime of the $QGP$. The increment in $\psi(2S)$ yield is then
\begin{equation}
\label{eq:increment}
\frac{N_{J/\psi}}{N_{\psi(2S)}}\Delta n_{1S\rightarrow 2S},
\end{equation}
where $\frac{N_{J/\psi}}{N_{\psi(2S)}}$ is the ratio of number of initial $J/\psi$ to $\psi(2S)$. As an estimate, we have taken $\frac{N_{J/\psi}}{N_{\psi(2S)}}$ to be equal to the ratio of the production cross-section, $\frac{\sigma^{NN}_{J/\psi}}{\sigma^{NN}_{\psi(2S)} } = \frac{1}{0.3}$~\cite{rituraj}. 


\section{Dissociation Processes}
\label{sec:diss}
We now look at modeling the dissociation processes. There can be multiple dissociation process, like, gluon induced dissociation, collisional damping and Chu and Matsui mechanism of suppression due to color screening.
\subsection{Gluon dissociation}
This mechanism is a very similar mechanism to the gluon induced enhancement derived in this work. The $\psi(2S)$ absorbs a gluon, gets converted to an octet state, and finally dissociates. The derivation of the gluon dissociation cross section, based on pNRQCD, is outlined in \cite{Nendzig}. Since gluon dissociation and the derivation of gluon enhancement are both based on the pNRQCD Lagrangian, and at leading order, it allows an apples to apples comparison between the two. This is discussed further in Sec.~\ref{sec:ALICEcomp}.
The cross section is given by
\begin{equation}
\begin{split}
\sigma_{diss,nl}(E_g) = \frac{\pi^2\alpha_s^u E_g}{N_c^2} \sqrt{\frac{m}{E_g + E_{nl}}} \\
\times \left ( \frac{l|J_{nl}^{q,l-1}|^2 + (l+1)J_{nl}^{q,l+1}|^2}{2l+1} \right ) 
\end{split}
\end{equation}
where $J_{nl}^{ql'}$ can be expressed using singlet and octet wave functions as:
\begin{equation}
J_{nl}^{ql'} = \int_0^\infty dr\,r \,g^*_{nl}(r)h_{ql'}(r)
\end{equation}
\\
In our simulations, we use identical values for all parameters between gluon induced enhancement and dissociation.
The center of mass energy of incoming $J/\psi$ varies from 3.26 GeV to 25.2 GeV. For incoming $\psi(2S)$, it varies from 3.83 GeV to 25.27 GeV. These values correspond to $p_T$ range from 1 GeV to 25 GeV.
As in the case of gluon induced enhancement, the cross section is then averaged over the modified gluon distribution $f_g(E_g,v_{rel},\theta)$, to obtain $\Gamma_{gdiss}$,
On the same lines as Eq.~\ref{eq:rateconstant}, we then obtain,
\begin{eqnarray}
\nonumber \Gamma_{gdiss}= \frac{1}{4\pi^2}\int^{1 GeV}_{E_{bind}} \frac{ E_g g_dT\,\sigma_{gdiss}}{v_{rel}\gamma} \\
\times \ln\Bigg [ \frac{e^{\frac{\gamma E_g}{T}(1 + v_{rel})} - 1}
{e^{\frac{2\gamma E_g v_{rel}}{T}} \left (e^{\frac{\gamma E_g}{T} (1- v_{rel})} - 1\right ) } \Bigg ]dE_g.
\end{eqnarray}

\subsection{Collisional damping}
Collisional damping is essentially the decay due to the imaginary part of the potential~\cite{Laine1} between the quark and antiquark.
\begin{equation} 
\label{eq:potential}
\begin{split}
	V(r,T) = \frac{\sigma_c}{m_D}(1 - e^{-m_D\,r}) \\ 
- \alpha_{eff} \left ( m_D + \frac{e^{-m_D\,r}}{r} \right ) \\
- i\alpha_{eff} T \int_0^\infty \frac{2\,z\,dz}{(1+z^2)^2} \left ( 1 - \frac{\sin(m_D\,r\,z)}{m_D\,r\,z} \right ),
\end{split}
\end{equation} 
where, $m_D$ is the Debye mass and is given by 
\begin{math}
	m_D = T\sqrt{4\pi\,\alpha_s^T \left (\frac{N_c}{3} + \frac{N_f}{6} \right ) }.
\end{math}

The collisional damping rate constant is then  given by~\cite{gans1}
\begin{equation}
\Gamma_{damp} = \int[\psi^\dagger_T \left [ Im(V(r,T))\right ] \psi_T]\,dr, 
\end{equation}
with $\psi_T$, being the singlet wavefunction at temperature T.
At different values of $p_T$, the singlet particle $\psi(2S)$, will effectively see a modified distribution of quarks and gluons colliding with it. This effect is captured by using an effective Temperature $T_{eff}$~\cite{Tefforig}, which then varies with $p_T$. 
\begin{equation}
T_{eff} = T\frac{\sqrt{(1 - v_{rel}^2)}}{1 - v_{rel}\cos(\theta)},
\end{equation}
where $\theta$ is the scattering angle.
Averaging over $d\Omega = \sin(\theta)d\theta d\phi$, we get
\begin{equation}
\label{eq:teff}
\langle T_{eff}\rangle = \frac{1}{2v_{rel}}T\sqrt{(1 - v_{rel}^2)}\ln\left (\frac{1 + v_{rel}}{1-v_{rel}}\right ),
\end{equation}
The singlet wavefunction $\psi$ is then determined at temperature $\langle T_{eff}\rangle$, to model the $p_T$ dependence of $\Gamma_{damp}$.  

The net dissociation constant due to collisional damping and gluon induced dissociation is given by 
\begin{equation}
	\Gamma_{total} = \Gamma_{damp} + \Gamma_{gdiss}
\end{equation}

\subsection{Color Screening}
In the context of color screening suppression, we model the mechanism of
suppression due to Debye color screening. We follow the formulation outlined
in~\cite{Madhu1,Madhu2}. 
To incorporate medium effects, we have used the relative velocity between the $\psi(2S)$ velocity, $\vec{v}_{\psi}$, and medium velocity, $\vec{v}_{med}$, in place of $\vec{v}_{\psi}$ in the color screening equation used in~\cite{Madhu1,Madhu2}, i.e.,
\begin{equation}
\label{eq:escape}
|\vec{r}_{\psi} + \vec{v}_{rel}t_F| < r_s,
\end{equation}
instead of 
\begin{equation}
\nonumber |\vec{r}_{\psi} + \vec{v}_{\psi}t_F| < r_s.
\end{equation}
Finally, we calculate suppression using the cooling law and pressure profile
discussed in ~\cite{Madhu2}. The transverse energy deposited per unit
rapidity $\frac{dE_{T}}{dy}$ and overlap area $A_T$ are required to obtained the
average pressure of the medium as the function of centrality  $N_{part}$ in
p$-$Pb collision. We use the experimental value of $\frac{dE_{T}}{d\eta}$ and obtained
the $dE_{T}/dy$ using the relation $dE_{T}/dy = 1.09\times dE_{T}/d\eta$. The
overlap area, $A_T$, has been calculated using the Monte Carlo Glauber model
within the framework of the ROOT software~\cite{ROOT}. The values of $N_{part}$
have been obtained from \cite{ALICEdata}, and $\frac{dE_T}{d\eta}$ from~\cite{etref}.

\section{Results and Discussions}
\label{sec:Results}
\subsection{Comparison with ATLAS data}
\label{sec:ATLAScomp}
We begin by discussing the effect of modified Bose-Einstein distribution $f_g(E_g,v_{rel},\theta)$ on the gluon density available for exciting $J/\psi(1S)$ to $\psi(2S)$.
\begin{figure}[h!]
\includegraphics[width = 70mm,height = 70mm]{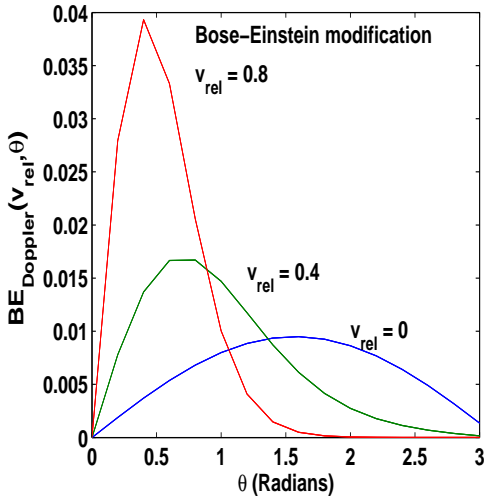}
\caption{Variation of $BE_{Doppler} = \sin(\theta)\times f_g(E_g,v_{rel},\theta)$, w.r.t. $\theta$, for $E_g = 0.8$ GeV, and various values of $v_{rel}$.}
\label{fig:doppler}
\end{figure}
We depict $BE_{Doppler} = \sin(\theta)\times f_g(E_g,v_{rel},\theta)$, w.r.t. $\theta$, for various values of $v_{rel}$ in Fig.~\ref{fig:doppler}. We have plotted $\sin(\theta)\times f_g(E_g,v_{rel},\theta)$, instead of $f_g(E_g,v_{rel},\theta)$, since the $\sin(\theta)$ term appears in the integrand of the expression for the rate constant, $\Gamma_{1S\rightarrow2S}$ (Eq.~\ref{eq:rateconstant}). 
In the blue shifted forward region, i.e., $\theta < \frac{\pi}{2}$, there is a substantial increase in the gluon density, as $v_{rel}$ increases. This gives an indication as to why gluo-excitation of $J/\psi$ to $\psi(2S)$ should increase with $p_T$. 

\begin{figure}[h!]
\includegraphics[width = 70mm,height = 70mm]{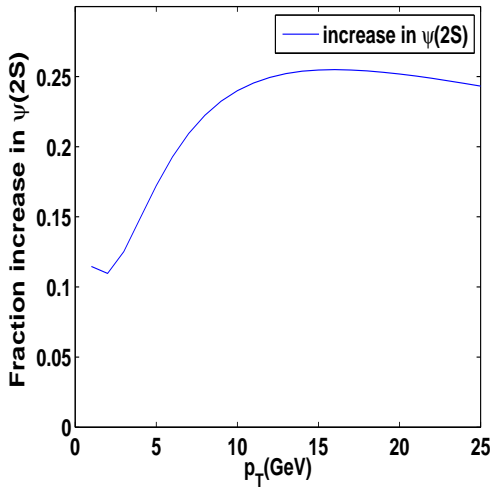}
\caption{The fractional increase in $\psi(2S)$ as a function of $p_T$, for medium velocity equal to $0.5c$.}
\label{fig:psi_inc}
\end{figure}

\begin{figure}[h!]
\includegraphics[width = 80mm,height = 80mm]{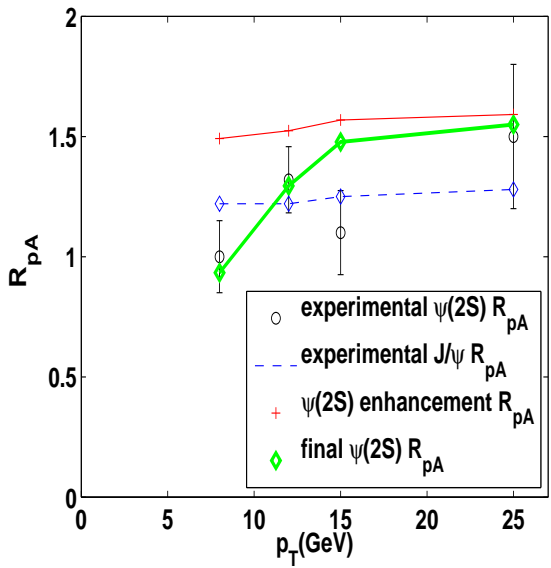}
\caption{Comparison of the final $\psi(2S)$ yield with the experimental ATLAS data~\cite{ATLAS} in p$-$Pb collision at $5.02$ TeV. The dashed blue line indicates prompt $J/\psi$ yield. $t_{QGP}$ = 3.0 fm for the most central bin.}
\label{fig:psi_yield_cs}
\end{figure}

We shall now discuss the simulation results.
The value of $E_{1S}$ and $E_{2S}$ are obtained by solving the Schr\"{o}dinger equation with the potential in Eq.~\ref{eq:potential}.
In the potential expression, we use, $N_f = 3$ = number of flavors and $\alpha_s^T = 0.10$. The value of $\sigma_{c}= 0.192~GeV^2$ for singlet.
The value of $\alpha_{eff} = \frac{4\alpha}{3}$ for singlet, 
and we have taken $\alpha = 0.27$. 
The Schr\"{o}dinger equation has been solved by taking a $10^4$ point logarithmically spaced finite spatial grid of size $9$~fm and solving the resulting matrix eigenvalue equations. 
We assume the critical temperature, $T_c$, to be $170$ MeV, and the charm quark mass equal to $1.275$ GeV. With these values, we determine $\Gamma_{1S\rightarrow 2S}$.
It is to be noted that the potential $V(r,m_D)$ would also modify the mass of $J/\psi$ and $\psi(2S)$ with temperature. 
The centrality and time dependence of the temperature of the QGP medium is taken as~\cite{gans1}, 
\begin{equation}
T(t) = T_c \frac{\left(\frac{dN_{ch}}{d\eta}/\frac{N_{part}}{2}\right)_{bin}^{1/3}} {\left(\frac{dN_{ch}}{d\eta}/\frac{N_{part}}{2}\right)_{bin0}^{1/3}} \left ( \frac{t_{QGP}}{t} \right )^{1/3}.
\end{equation}
Here, $bin0$ refers to a reference bin, taken as the most central bin. The variable, $bin$, corresponds to the index varying from the most central to the most peripheral bin.
The simulation has been done for $t_{QGP}$ = $3.0$ fm for the most central bin. The value of $t_0$ used is $0.6$ fm.
The values for $N_{part}$ and $\frac{dN_{ch}}{d\eta}$ have been obtained from~\cite{ALICEdata}. 
The analytical form of the correlation function $\langle 1S|r|O(l=1)\rangle$ has been taken from~\cite{nora1}; 
\begin{eqnarray}
\nonumber |\langle 1S|r|O(l=1)\rangle| = \\
\nonumber \sqrt\frac{512\pi^2\rho(\rho + 2)^2 a_0^6 \left (1 + \frac{\rho^2}{a_0^2 q^2} \right ) e^{\frac{4\rho}{a_0q} tan^{-1}(a_0q)}} 
{\left ( e^{\frac{2\pi \rho}{a_0 q}} - 1 \right ) \left (1 + a_0^2q^2)^6 \right )}\\
\end{eqnarray}
where $\rho = \frac{1}{N_c^2 - 1}$ and $a_0$ is the Bohr radius.
Similarly, the correlation function $\langle 2S|r|O(l=1)\rangle$ has been inferred from~\cite{nora1};
\begin{eqnarray}
\nonumber |\langle 2S|r|O(l=1)\rangle| = \\
\nonumber \sqrt \Bigg \{ \frac{3 \pi^2 2^{12} \rho }{(m_c^2 E^4 (1 + 4a0^2 q^2)^8 q )} \\
\nonumber \left (2 E_{2S} (2\rho^2 + 5\rho + 3) + q(\rho + 2) \right )^2 \\
\nonumber \left ( (2\,a_0\,q)^2 + 4 \rho^2 \right ) \\
\times e^{\frac{8\rho}{2a_0\,q}tan^{-1}(2a_0\,q)} \left [e^{\frac{4\pi \rho}{2a_0\,q}} - 1\right ]^{-1} \Bigg \}
\end{eqnarray}
Figure~\ref{fig:psi_inc} depicts the fractional increase in $\psi(2S)$ yield as a function of $p_T$. We see that this increases with $p_T$. 
Apart from the transition of $J/\psi(1S)$ to $\psi(2S)$, there would be multiple factors contributing to the $\psi(2S)$ yield. These include CNM effects, suppression effects due to Debye color screening, gluon induced dissociation etc. Absorption due to CNM effects can be expected to be negligible at the LHC~\cite{vogt}. The other CNM effects, namely Cronin effect and shadowing, being initial state effects, would be expected to be the same between $J/\psi$ and $\psi(2S)$. Hence, we take the $J/\psi$ $R_{pA}$ as a measure of CNM. 
For ease of reference, the $J/\psi$ $R_{pA}$ in the p$-$Pb collision is also plotted in Fig.~\ref{fig:psi_yield_cs}.

\begin{figure}[h!]
\includegraphics[width = 80mm,height = 80mm]{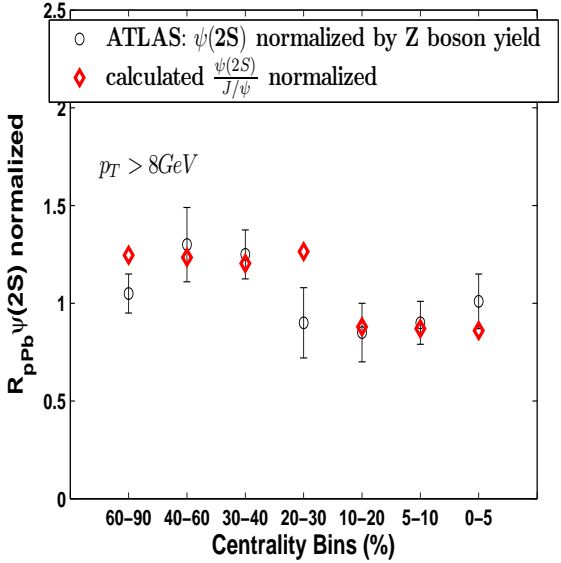}
\caption{Comparison with the ATLAS~\cite{new_ATLAS} experimental $\psi(2S)$ yield, which is normalized by the Z boson yield.} 
\label{fig:new_atlas_z}
\end{figure}

\begin{figure}[h!]
\includegraphics[width = 80mm,height = 80mm]{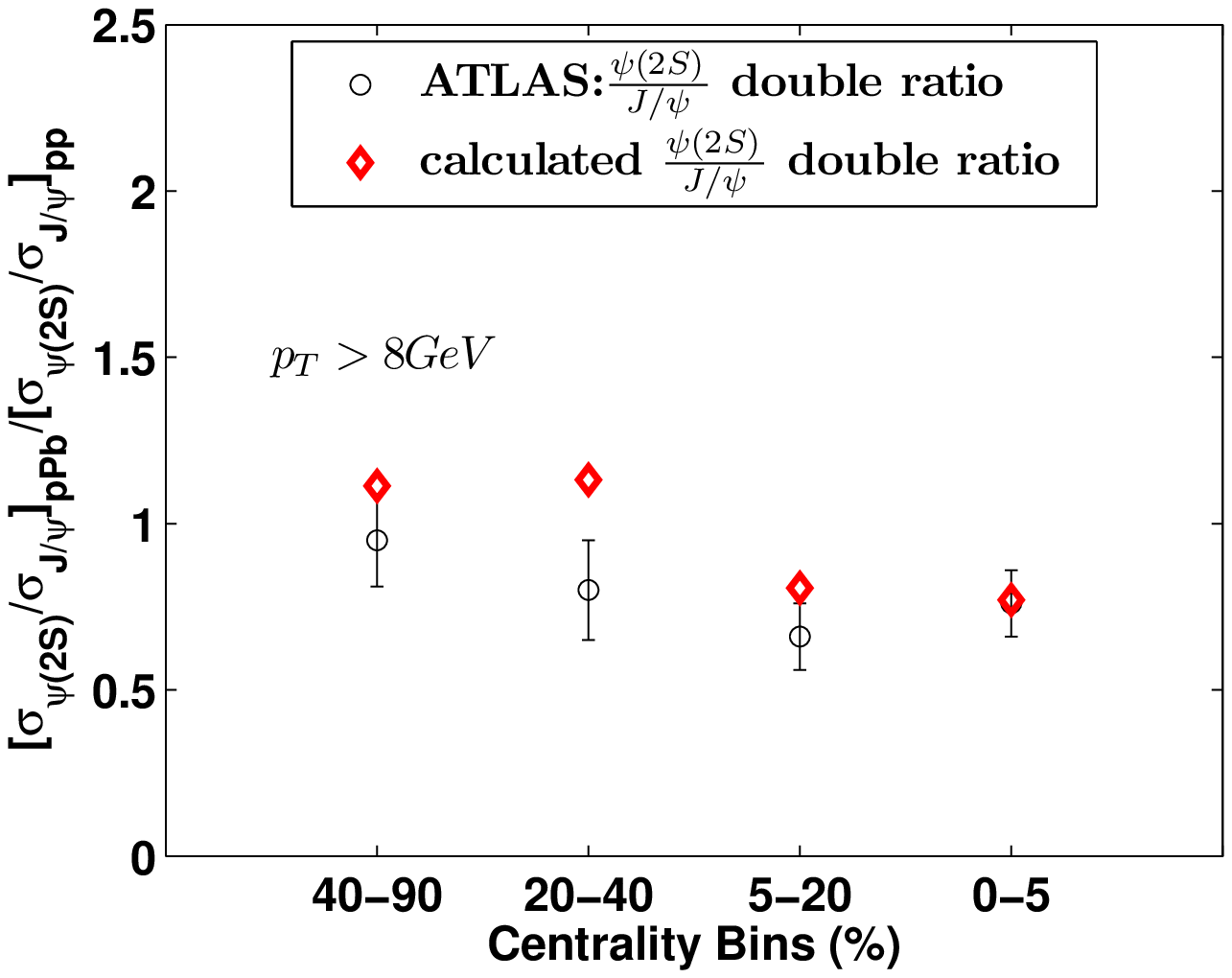}
\caption{Comparison with the ATLAS~\cite{new_ATLAS} experimental $\psi(2S)$ double ratio.} 
\label{fig:new_atlas_dratio}
\end{figure}

\begin{figure}[h!]
\includegraphics[width = 80mm,height = 80mm]{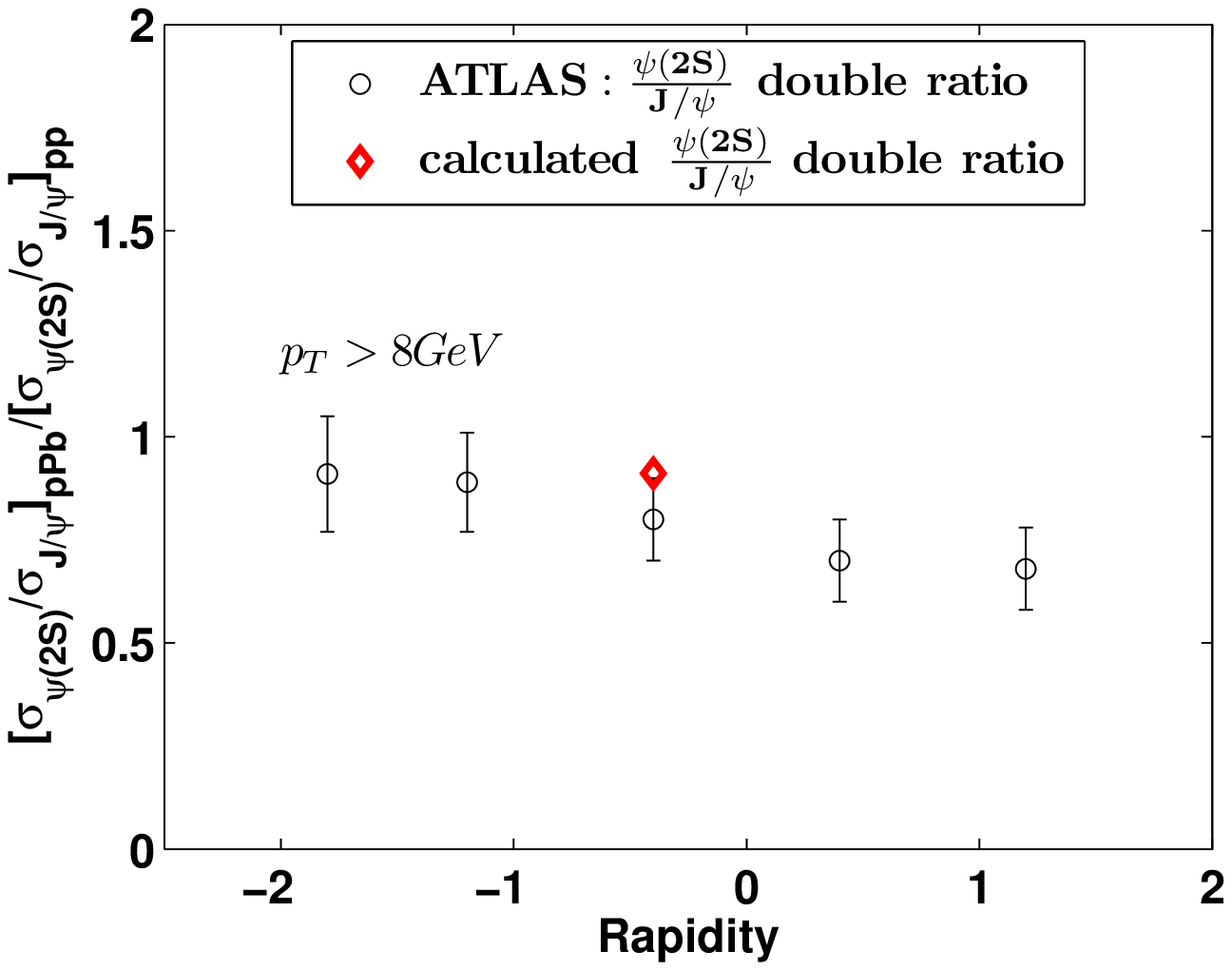}
\caption{Comparison with the ATLAS~\cite{new_ATLAS} experimental $\psi(2S)$ double ratio.} 
\label{fig:new_atlas_rapidity}
\end{figure}

In Fig.~\ref{fig:psi_yield_cs}, we compare the $\psi(2S)$ yield after gluo-excitation (red curve) with the ATLAS experimental data~\cite{ATLAS}. 
We shall later compare with the other ATLAS data~\cite{new_ATLAS}, in Figs.~\ref{fig:new_atlas_z},~\ref{fig:new_atlas_dratio} and \ref{fig:new_atlas_rapidity}.
In Fig.~\ref{fig:psi_yield_cs}, the $\psi(2S)$ yield includes the $J/\psi(1S)$ yield as a measure of CNM. 
The $p_T$ dependent simulation data has been obtained by weighing each centrality bin with the corresponding value of $N_{coll}$ and then averaging. The $N_{coll}$ values have been obtained from ~\cite{ALICEdata}. The centrality bins are identical to those in ATLAS~\cite{ATLAS, new_ATLAS}.
The regions in the plot, where the experimental yield of $\psi(2S)$ is close to unity, is a probable indication of suppression effects. 

To include suppression, we model the mechanism of suppression due to gluon induced dissociation, collisional damping and Debye color screening, described in Sec.~\ref{sec:diss}.

There is a very important difference between gluo-excitation and gluo-suppression. In gluo-excitation, the intermediate octet state can emit a gluon and transition to the $\psi(2S)$ bound state even in the hadronic phase after the QGP has ended. In fact, it may be energetically more favorable to end up as a $\psi(2S)$ bound state, rather than completely dissociate into the constituent $c\bar{c}$ pair.  
However for gluo-dissociation, its unlikely that the excited intermediate octet state would dissociate into the constituent $c\bar{c}$ pair in the hadronic phase. 
The light quarks and anti-quarks, which have been (color) screening the $c\bar{c}$ pair from each other, disappear with the QGP transitioning to the hadronic phase.
The hadronic phase does not contain naked light quarks to bind with $c$ or $\bar{c}$ to form $D$ mesons either.
Thus the gluon induced dissociation needs to complete within the QGP lifetime itself. 
The $\psi(2S)$ binding energy is about $0.05$ GeV in vacuum. But if a gluon of energy $0.05$ GeV interacts with $\psi(2S)$, the intermediate octet state may evolve for about $1/0.05$ GeV $\approx$ $3.95$ fm, before it fully dissociates. But, by this time, the QGP lifetime gets over, and hadronic phase comes into existence. As a result, the excited octet state may no longer dissociate. Rather, it is more likely to emit a gluon, and end up as a bound state.
To ensure that the dissociation takes place within the QGP lifetime itself,
a much higher gluon energy is required to dissociate $\psi(2S)$.
We take the minimum energy of gluon to cause suppression to be $\frac{\gamma}{t_{QGP}}$.
This curtails the dissociation of $\psi(2S)$ to a significant extent.
This phenomenon need not be significant in central and mid-central Pb-Pb collisions, where QGP lifetimes are much higher.

The final $\psi(2S)$ yield (green curve) after including the suppression due to all the mechanisms is depicted in Fig.~\ref{fig:psi_yield_cs}. The $p_T$ dependence of suppression depicted in Fig.~\ref{fig:psi_yield_cs}, is after taking into account the effect of the radial medium velocity, $\vec{v}_{med}$. 
The medium velocity is taken as $0.5$ based on~\cite{medvel}.
The absence of a more precise modeling of the QGP expansion, utilizing a $3+1$-hydrodynamical model, is a limitation of this work.
It can be seen here that with suppression included, the experimental data is better captured. The results show that the $\psi(2S)$ suppression is relatively high at low $p_T$, while gluo-excitation of $J/\psi(1S)$ to $\psi(2S)$ is high at high $p_T$. 
This gets highlighted in Fig.~\ref{fig:en_gdiss_comp}, which is discussed further in Sec.~\ref{sec:ALICEcomp}.
A note on the effect of $J/\psi$ suppression is due at this point. We expect the temperature of the QGP, if it is formed, to be not very high. The $J/\psi$ dissociation temperature is about $381$ MeV, which would almost always be above the QGP temperature formed in p$-$Pb collisions. Thus, $J/\psi$ will not undergo any suppression due to Debye color screening. 
Gluon induced suppression and suppression due to collisional damping for $J/\psi$, is depicted in Fig.~\ref{fig:en_gdiss_comp}, which indicate this to be small as well.

At this point, it is worthwhile to discuss about what happens to a $p_T$ integrated $\psi(2S)$ yield. Since most of the $\psi(2S)$ production would be expected to be in the low $p_T$ region, the net $\psi(2S)$ yield would be dominated by what occurs in the low $p_T$ region, which is suppression. 
We discuss this in detail in Sec.~\ref{sec:ALICEcomp}.
In peripheral p$-$Pb collisions, where both suppression and gluo-excitation is expected to be low, there would be no enhancement or suppression w.r.t. $J/\psi$, if QGP exists in such a collisions. However, the QGP formation probability seems to be quite small in a peripheral collisions.     
A question that arises here is whether the enhancement phenomenon would be seen in Pb$-$Pb collision. In a Pb$-$Pb collision, the temperatures of the medium are much higher, and can go upto $400$ or $500$ MeV. At temperatures above $190$ MeV, $\psi(2S)$ cannot exist. Any $\psi(2S)$ present would only dissociate. The gluo-excitation to $\psi(2S)$ cannot happen during this period of the QGP. During the short time, when the temperature of the QGP reduces to below $190$ MeV, this phenomenon may happen, but may get overshadowed by the suppression phenomenon that has been happening throughout the lifetime of the QGP. However, in the extreme peripheral collisions, where the temperatures are much lower during the QGP lifetime, there could be a possibility of gluon induced $\psi(2S)$ enhancement at high $p_T$.  

We now compare with the ATLAS data \cite{new_ATLAS} in Figs.~\ref{fig:new_atlas_z},~\ref{fig:new_atlas_dratio} and \ref{fig:new_atlas_rapidity}. Integrated $p_T$ yield ($p_T > 8$ GeV) is used to compare with experimental data. 
The $\psi(2S)$ values around $8$ GeV, play the dominant role as $\psi(2S)$ production decreases significantly with increasing $p_T$~\cite{new_ATLAS}.
In Fig.~\ref{fig:new_atlas_z}, we plot the calculated $\frac{\psi(2S)_{zb}}{\sum_{bins} \psi(2S)_{zb}}$ data, where $\psi(2S)_{zb} = \frac{\psi(2S)_{bin}N_{coll}}{Z_{bin}}$. 
The Z boson data, $Z_{bin}$, obtained from~\cite{Zboson}, is the yield after normalization with $N_{coll}$, and without any Glauber Gribov color fluctuations. 
Figure~\ref{fig:new_atlas_dratio} compares the $\frac{\psi(2S)}{J/\psi}$ double ratio. Our simulation overestimates the double ratio somewhat, but it is able to capture the trend. 
Finally, in Fig.~\ref{fig:new_atlas_rapidity}, we compare our simulation results with the ATLAS rapidity dependence data in \cite{new_ATLAS}. Our result at mid-rapidity is just about touching the error bar.

\subsection{Comparison with ALICE data}
\label{sec:ALICEcomp}
Unlike ATLAS data, ALICE data \cite{alicepsi} shows suppression. This suppression is observed at low $p_T$ range. Therefore, in order to compare with the ALICE data, we extend our simulation results to low $p_T$ region.

\begin{figure}[h!]
\includegraphics[width = 80mm,height = 80mm]{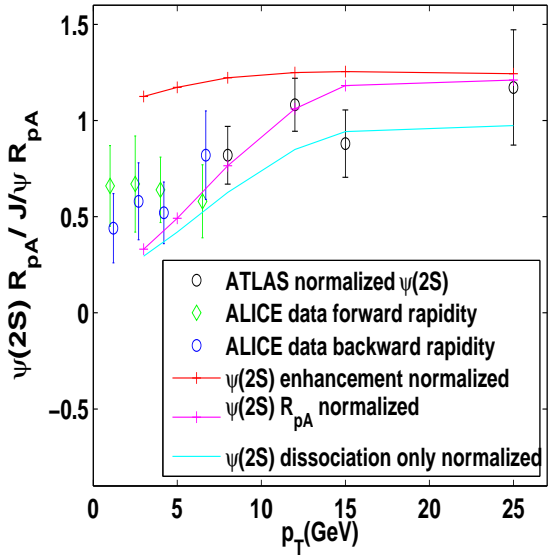}
\caption{Comparison of simulated results with both ATLAS~\cite{ATLAS} and ALICE~\cite{alicepsi} data.}
\label{fig:ATLAS_ALICE}
\end{figure}

Fig.~\ref{fig:ATLAS_ALICE} show the simulated data compared with both ATLAS~\cite{ATLAS} and ALICE~\cite{alicepsi} data.
The ALICE data is the double ratio of $\tfrac{\sigma_{2S}}{\sigma_{1S}}$. In order to keep both the ALICE and ATLAS data on the same footing, we normalize the ATLAS $\psi(2S)$ $R_{pA}$ data by dividing with the baseline $R_{pA}$ for $J/\psi$. 
This also enables us to ignore CNM effects, as the CNM effects of Cronin effect and shadowing are initial state effects, and expected to be the same in both, $J/\psi$ and $\psi(2S)$.
We see that the simulation data is able to capture the trend of both ALICE~\cite{alicepsi} and ATLAS~\cite{ATLAS} data simultaneously. 
We need to note here that the ALICE data is at forward and backward rapidity, while the ATLAS data is at mid rapidity. Our simulation hydrodynamical model also essentially models the mid rapidity region. Hence, we do not expect our simulation to identically match ALICE data, but only capture the suppression trend.  
However, it is of significance that our model is able to simultaneously capture the enhancement at ATLAS at high $p_T$, and suppression at ALICE at low $p_T$.

\begin{figure}[h!]
\includegraphics[width = 80mm,height = 80mm]{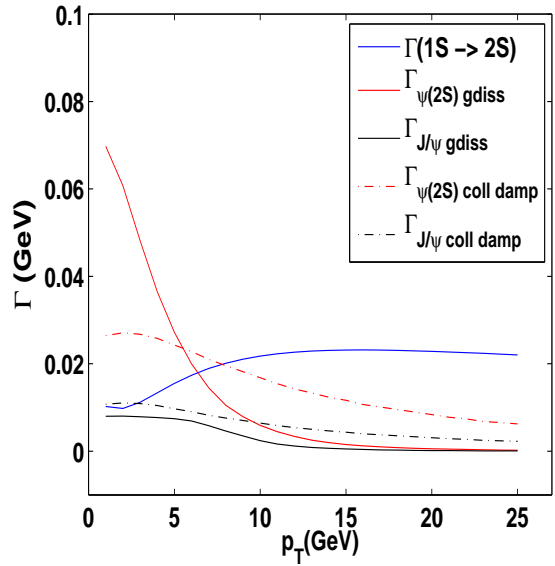}
\caption{Comparison of $\Gamma_{diss}$ due to gluodissociation and collisional damping with $\Gamma_{1S \rightarrow 2S}$. Depicted $\Gamma_{diss}$ due to gluodissociation is averaged over all bins.}
\label{fig:en_gdiss_comp}
\end{figure}

We try to understand the above result by first comparing the similar mechanisms of gluon induced dissociation of $\psi(2S)$ and enhancement of $J/\psi$ to $\psi(2S)$ in Fig.~\ref{fig:en_gdiss_comp}.  
We have used identical values of all the parameters like coupling constant etc. to calculate $\Gamma_{gdiss}$ and $\Gamma_{1S \rightarrow 2S}$.
We see that dissociation is higher at low $p_T$, while enhancement is higher at high $p_T$. The binding energy of $\psi(2S)$ in vacuum is about $0.05$ GeV which is much lower than the energy gap between $J/\psi$ and $\psi(2S)$. 
Inspite of the low binding energy, $\psi(2S)$ dissociation is subdued since much higher gluon energy is required for $\psi(2S)$ to dissociate within the QGP, as discussed earlier in Sec.~\ref{sec:ATLAScomp}.
 At high $p_T$, when the $\psi(2S)$ absorbs a gluon, the evolution of the meson will happen very slowly due to Lorentz time dilation. Hence, significantly more gluon energy is required to dissociate it faster, while the meson is still within the QGP. As mentioned earlier, the minimum energy of gluon required would be $\frac{\gamma}{t_{QGP}}$, with $\gamma$ being the Lorentz dilation factor. This results in a drastic decrease in  $\Gamma_{gdiss}$ at high $p_T$.

	We have also observed that the other suppression effects like collisional damping and Debye color screening are also higher at low $p_T$. In the case of collisional damping, the value of $T_{eff}$, given by Eq.~\ref{eq:teff}, is lower at high $p_T$, leading to lower suppression at high $p_T$. The suppression due to collisional damping is likely to have similar restrictions as gluon induced dissociation due to small QGP lifetime. We plan to quantitatively model such effects in future.

   The enhancement of $\psi(2S)$ is further amplified by the factor $\frac{N_{J/\psi}}{N_{\psi(2S)}}$ in Eq.~\ref{eq:increment}. Every fraction of $J/\psi$ converted to $\psi(2S)$, gives rise to $\frac{N_{J/\psi}}{N_{\psi(2S)}}$ fraction increase in $\psi(2S)$. This factor can lead to even higher enhancement of $\psi(2S)$ than dissociation of $\psi(2S)$. 
The net $\psi(2S)$ $R_{pA}$, seems to be a balance of the enhancement and dissociation process, and tilts in the favor of which ever is higher.
	As a final remark, even if the enhancement of $\psi(2S)$, were not to dominate over suppression of $\psi(2S)$, the phenomenon of $\psi(2S)$ enhancement may not be ignorable for capturing $\psi(2S)$ yield at high $p_T$.
This is seen from Fig.~\ref{fig:ATLAS_ALICE}, where the normalized $\psi(2S)$ $R_{pA}$ double ratio (cyan curve), with only dissociation modeled, somewhat underestimates the $\psi(2S)$ $R_{pA}$ double ratio. 

\begin{figure}[h!]
\includegraphics[width = 80mm,height = 80mm]{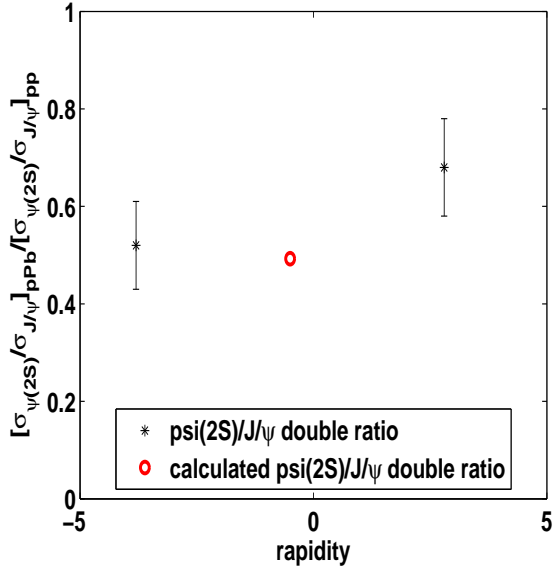}
\caption{Comparison with ALICE  rapidity data for the double ratio $\psi(2S)/J/\psi]_{pPb}]/[\psi(2S)/J/\psi]_{pp}$~\cite{alicepsi}. 
}
\label{fig:alicerapidity}
\end{figure}

Figure~\ref{fig:en_gdiss_comp} also shows that $J/\psi$ dissociation is small. This seems to be in agreement with other literature~\cite{medvel}, where CNM effects have been mainly used to analyze $J/\psi$ $R_{pA}$. 
Finally, we explore the nature of $\psi(2S)$ modification for $p_T$ integrated data.
\begin{figure}[h!]
\includegraphics[width = 80mm,height = 80mm]{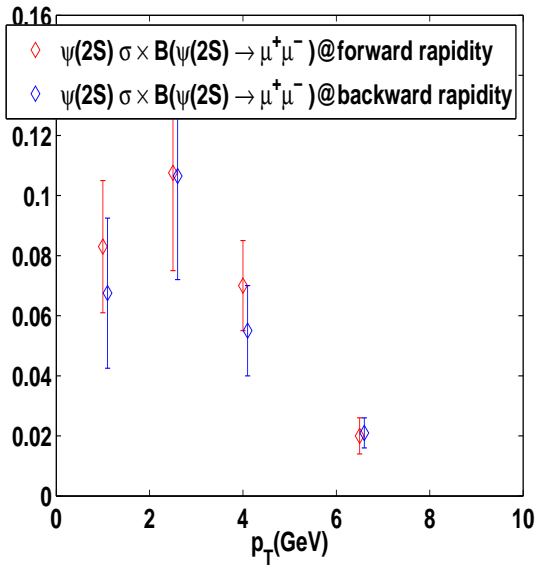}
\caption{CMS~\cite{cmsjpsi} data for $\psi(2S)$ distribution.}
\label{fig:psi2S_prod_pt}
\end{figure}
Figure ~\ref{fig:psi2S_prod_pt} shows CMS experimental data on $\psi(2S)$ distribution as a function of $p_T$~\cite{cmsjpsi}. It can be seen that most of $\psi(2S)$ is concentrated at low $p_T$ region.
Thus for any $p_T$ integrated evaluation of $\psi(2S)$ yield, the modification of $\psi(2S)$ at low $p_T$, which is suppression, is expected to dominate.  
Figure~\ref{fig:alicerapidity} shows that the $p_T$ integrated value at mid rapidity is suppression and not enhancement. It is to be noted that the experimental data~\cite{alicepsi} indicates suppression and not enhancement at forward and backward rapidity.

\section{Conclusions}
\label{sec:conclusion}
In conclusion, we have attempted to explain the transverse momentum dependence of $\psi(2S)$ suppression data observed in p$-$Pb collision at the LHC energy, over a wide span of transverse momentum. We have found a differential enhancement of $\psi(2S)$ w.r.t. $J/\psi(1S)$ at higher transverse momentum using gluon induced $1S$ to $2S$ transition approach, which to a significant extent agrees with the preliminary ATLAS experimental data. 
We have also included the effect of suppression via the gluo-dissociation and collisional damping mechanisms and the Chu and Matsui mechanism of Debye color screening. 
The combined result seems to corroborate better with the experimental ATLAS data. Our simulations results also corroborate with the trend in ALICE results for $\psi(2S)$ for both $p_T$ and rapidity dependence, where suppression is the dominant phenomenon. We do not see any $J/\psi$ dissociation, which augurs well with the fact that CNM effects have been able to explain $J/\psi$ $R_{pA}$ in literature. We expect that the enhancement of $J/\psi(1S)$ to $\psi(2S)$, as an increasing function of $p_T$, if confirmed experimentally, can be one definitive evidence for the presence of QGP in p$-$Pb collision. 
Even if there is no net enhancement at high $p_T$,  $\psi(2S)$ enhancement seems to be required in order to predict the $\psi(2S)$ $R_{pA}$, especially at high $p_T$.
After submission of this manuscript, we became aware that the data~\cite{ATLAS} has been superseded by~\cite{new_ATLAS}. 
We have, however, retained the superseded ATLAS data~\cite{ATLAS}, as it is likely that there is a value addition in showing that our simulation results corroborate with the superseded ATLAS data. 
\section{Acknowledgments}
M. Mishra is grateful to the Department of Science and Technology (DST), New Delhi for financial assistance. Captain R. Singh is grateful to BITS - Pilani, Pilani for the financial assistance.

\end{document}